\documentclass[lettersize,journal]{IEEEtran}
\usepackage{amsmath,amsfonts, amssymb}
\usepackage{array}
\usepackage[caption=false,font=normalsize,labelfont=sf,textfont=sf]{subfig}
\usepackage{textcomp}
\usepackage{stfloats}
\usepackage{url}
\usepackage{verbatim}
\usepackage{graphicx}
\usepackage{tabularx}
\usepackage{booktabs}
\usepackage{algorithm}
\usepackage[noend]{algpseudocode}
\usepackage{multirow} 
\usepackage{makecell}
\usepackage{lipsum}
\usepackage{scalerel}
\usepackage{comment}
\usepackage{enumitem}
\usepackage{todonotes}
\usepackage{bm}
\setlist[itemize]{noitemsep, nolistsep}
\setlist[enumerate]{noitemsep, nolistsep}
\usepackage{xcolor}  
\def\BibTeX{{\rm B\kern-.05em{\sc i\kern-.025em b}\kern-.08em
    T\kern-.1667em\lower.7ex\hbox{E}\kern-.125emX}}
\usepackage{balance}

\usepackage{pifont}

\usepackage[utf8]{inputenc}
\usepackage{kotex}
\usepackage{xspace}
\usepackage{calc}

\newcommand{\rev}[1]{{\color{blue}#1}}

\definecolor{jumincolor}{RGB}{0,200,0}

\newcommand{\OHprac}{\ensuremath{\mathrm{OH}_{\mathrm{prac}}}\xspace}
\newcommand{\actcmd}{\texttt{ACT}\xspace}
\newcommand{\precmd}{\texttt{PRE}\xspace}
\newcommand{\rdcmd}{\texttt{RD}\xspace}
\newcommand{\wrcmd}{\texttt{WR}\xspace}

\newcommand{\tRC}{$t_{\mathrm{RC}}$\xspace}
\newcommand{\tRCD}{$t_{\mathrm{RCD}}$\xspace}
\newcommand{\tRAS}{$t_{\mathrm{RAS}}$\xspace}
\newcommand{\tRP}{$t_{\mathrm{RP}}$\xspace}
\newcommand{\tWR}{$t_{\mathrm{WR}}$\xspace}
\newcommand{\tRTP}{$t_{\mathrm{RTP}}$\xspace}
\newcommand{\tCCDL}{$t_{\mathrm{CCDL}}$\xspace}
\newcommand{\tCL}{$t_{\mathrm{CL}}$\xspace}

\newcommand{\tRASbold}{\ensuremath{\bm{t_{\mathrm{RAS}}}}\xspace}
\newcommand{\tRPbold}{\ensuremath{\bm{t_{\mathrm{RP}}}}\xspace}
\newcommand{\tRCbold}{\ensuremath{\bm{t_{\mathrm{RC}}}}\xspace}
\newcommand{\tRTPbold}{\ensuremath{\bm{t_{\mathrm{RTP}}}}\xspace}
\newcommand{\tWRbold}{\ensuremath{\bm{t_{\mathrm{WR}}}}\xspace}

\begin{document}

\title{Per-Row Activation Counting on Real Hardware: Demystifying Performance Overheads}

\author{
Jumin~Kim, 
Seungmin~Baek,
Minbok~Wi,
Hwayong~Nam,
Michael~Jaemin~Kim,
Sukhan~Lee,
\\Kyomin~Sohn, 
and~Jung~Ho~Ahn,~\IEEEmembership{Senior Member, IEEE}
\\
\thanks{This work was partly supported by Samsung Electronics Co., Ltd (IO241204-11396-01) and IITP (RS-2021-II211343 and RS-2024-00402898). We appreciate the inspiring discussion with professor Moin Qureshi regarding the impact of PRAC on real machines and row-buffer management policies.}
\thanks{Jumin~Kim, Seungmin~Baek, Minbok~Wi, Hwayong~Nam, Michael Jaemin Kim and Jung Ho Ahn are with Seoul National University, Seoul 08826, South Korea. E-mail: \{tkfkaskan1, qortmdalss, homakaka, nhy4916, michael604, gajh\}@snu.ac.kr.}
\thanks{Sukhan~Lee and Kyomin~Sohn are with Samsung Electronics Corporation, Hwaseong-si, Gyeonggi-do 18448, South Korea. E-mail: \{sh1026.lee, kyomin.sohn\}@samsung.com.}
}

\maketitle
\begin{abstract}
Per-Row Activation Counting (PRAC), a DRAM read disturbance mitigation method, modifies key DRAM timing parameters, reportedly causing significant performance overheads in simulator-based studies.
However, given known discrepancies between simulators and real hardware, real-machine experiments are vital for accurate PRAC performance estimation.
We present the first real-machine performance analysis of PRAC.
After verifying timing modifications on the latest CPUs using microbenchmarks, our analysis shows that PRAC's average and maximum overheads are just 1.06\% and 3.28\% for the SPEC CPU2017 workloads---up to 9.15\(\times\) lower than simulator-based reports.
Further, we show that the close page policy minimizes this overhead by effectively hiding the elongated DRAM row precharge operations due to PRAC from the critical path.

\end{abstract}


\section{Introduction}
\label{sec:1_Introduction}

\IEEEPARstart{P}{er}-Row Activation Counting (PRAC), a newly introduced feature to prevent DRAM read disturbance errors, modifies key DRAM timing parameters while providing full tracking capability of DRAM activations.
Under PRAC, each DRAM row carries an activation counter alongside its data.
To update this counter, DRAM performs a read-modify-write operation at every precharge, leading to the modification of major DRAM timing parameters.
These timing changes inevitably affect the performance of general workloads.

While previous studies have reported significant performance overhead from PRAC timing parameters (\OHprac), their analysis is limited to simulator-based evaluation~\cite{hpca-2025-chronus, hpca-2025-autorfm, isca-2025-mopac, hpca-2025-qprac}.
In particular, Chronus~\cite{hpca-2025-chronus} demonstrated that PRAC incurs an average overhead of 9.7\%, with worst-case penalties reaching up to 13.4\% in memory-intensive workloads.\footnote{
The authors of the Chronus paper~\cite{hpca-2025-chronus} acknowledged the bug in their simulator and updated all their results in the arXiv version of the Chronus paper~\cite{chronus-arxiv-2025}. The revised results correct the Chronus paper's performance numbers but do not change the Chronus paper's conclusions.
Note that the bug fix also appeared in the artifact evaluation of QPRAC~\cite{qprac-ae}.}
In response to these findings, architectural solutions have been proposed to reduce this overhead~\cite{hpca-2025-chronus, isca-2025-mopac}.

Simulator-based evaluations might fall short of accurately modeling the performance characteristics of real-world memory systems when a simulator is not properly configured~\cite{micro-2024-mess,luo2025cleaningmess}, inherent microarchitectural complexity causes discrepancy between what is modeled in a simulator and the corresponding hardware (e.g., hardware prefetcher~\cite{taco-prefetch}), or a simulator contains an undiscovered bug.
These highlight the need for evaluation on real machines to accurately quantify \OHprac. 

In this paper, we apply PRAC timing parameters to the latest Intel processors and measure \OHprac. 
We verify that each modified DRAM timing parameter is correctly applied to the real machines using a microbenchmark.
Our results show that PRAC timing parameters cause an average performance degradation of only 1.06\,\% on SPEC~CPU2017~\cite{spec_cpu_2017} workloads, which is greatly lower than the overheads reported by simulator-based work.
Further, we analyze hardware events correlated with \OHprac and identify a page policy that can minimize it.
To the best of our knowledge, this is the first study to evaluate and analyze \OHprac on real systems.

\section{PRAC and DRAM Timing}
\label{sec:2_Background}
\subsection{DRAM Command Sequence and Timing Constraints}
\label{subsec:2_1_dram}

\begin{figure}[!tb]
  \center
  \includegraphics[width=0.99\columnwidth]{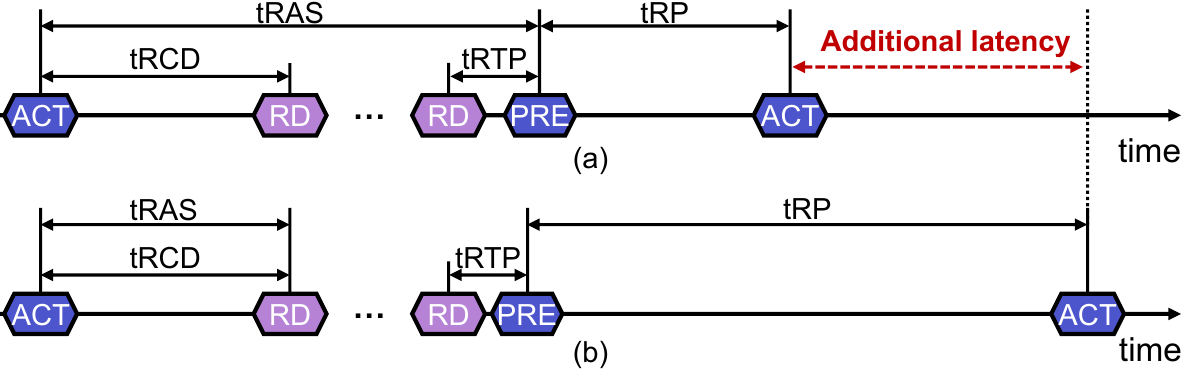}
  \caption{Timing diagram of DRAM command sequences under the systems DDR5 (a) without PRAC and (b) with PRAC.}
  \label{fig:command_sequence}
\end{figure}

\begin{table}[]
  \caption{Timing parameters for DDR5-4800 Default and PRAC}
  \label{tab:dram_timings}
  \centering
  \begin{tabularx}{\columnwidth}{@{} p{0.6cm} p{5.4cm} r r @{}}
    \toprule
    Timing & Description & Default & PRAC \\
    \midrule
    \tRASbold      & Min. time from \texttt{ACT} to \texttt{PRE}            &  32\,ns   & 16\,ns  \\
    \tRPbold       & Min. time from \texttt{PRE} to \texttt{ACT}            &  16\,ns   & 36\,ns  \\
    \tRCbold       & Min. time between consecutive \texttt{ACT}s            &  48\,ns   & 52\,ns  \\
    \tRTPbold      & Min. time from \texttt{RD} to \texttt{PRE}             & 7.5\,ns   &  5\,ns  \\
    \tWRbold       & Min. time from the end of write burst to \texttt{PRE}  &  30\,ns   & 10\,ns  \\
    \tRCD      & Min. time from \texttt{ACT} to \texttt{RD}/\texttt{WR} &  16\,ns   & 16\,ns  \\
    \tCL       & Min. time from \texttt{RD} to first data               &  16\,ns   & 16\,ns  \\
    \bottomrule
  \end{tabularx}
  \vspace{-0.1in}
\end{table}

DRAM access involves a memory controller (MC) issuing commands to a bank, a 2D array of rows and columns.
The MC issues a predefined sequence of DRAM commands while adhering to DRAM timing constraints (Figure~\ref{fig:command_sequence}).
An activate (\actcmd) command opens a row;
after $t_{\mathrm{RCD}}$, a read (\rdcmd) or write (\wrcmd) command can be issued to the target column.
Data for \rdcmd or \wrcmd is accessed from the requested column of the activated row through I/O pins, taking \tCL.
To issue consecutive \texttt{RD}s or \texttt{WR}s to the same row, the \tCCDL timing must be met.
To access a different row in the same bank, a precharge (\precmd) command can be issued after \tRAS, the minimum gap between \actcmd and \precmd.
\precmd closes a row over a period defined by \tRP.

Each DRAM bank has a row buffer that stores the currently activated row. 
If a subsequent access targets the same, already activated row, it results in a row-buffer hit.
This allows the access to proceed without issuing a new \precmd or \actcmd; only \tCL is required in DRAM.
In contrast, a row-buffer miss occurs when a different row in the same bank is accessed, requiring a \precmd followed by a new \actcmd, causing additional latency
(in total \tRP + \tRCD + \tCL).
When no row is activated, the bank is in the row-buffer empty state.
Then, an access requires only activating the target row, resulting in a latency of \tRCD + \tCL.

The frequency of row-buffer hits and misses is largely affected by the MC's page (row-buffer) management policy~\cite{intel-2024-pagepolicy}. 
An open page policy (\emph{open}) keeps rows open for a certain time anticipating row-buffer hits, improving performance for workloads with high spatial locality.
In contrast, a close-page policy (\emph{close}) aggressively precharges a row after each access, returns the bank to the row-buffer empty state, enabling faster subsequent activations.
Modern systems typically implement an adaptive page policy (\emph{adaptive}), which dynamically adjusts between \emph{open} and \emph{close} based on observed access patterns.

\begin{table}[t]
  \centering
  \caption{Summary of Prior studies on PRAC Timing Overhead}
  \vspace{-0.05in}
  \label{tab:prior_works_transposed}
  \footnotesize
  \setlength{\tabcolsep}{1pt} 
  \begin{tabularx}{0.95\linewidth}{@{}l|*{3}{>{\centering\arraybackslash}X}@{}}
    \toprule
      & Chronus~\cite{hpca-2025-chronus}
      & AutoRFM~\cite{hpca-2025-autorfm}
      & MoPAC~\cite{isca-2025-mopac} \\
    \midrule
    Simulator
      & Ramulator 2.0~\cite{ramulator2}
      & Memsim~\cite{memsim}
      & DRAMsim3~\cite{dramsim3} \\
    \midrule
    \makecell[l]{\OHprac}
      & \makecell[c]{Avg.\ 9.7\%\\Max.\ 13.4\%}
      & Avg.\ \(\ge\)4\%
      & \makecell[c]{Avg.\ 10\%\\Max.\ \(>15\%\)} \\
    \bottomrule
  \end{tabularx}
  \vspace{-0.14in}
\end{table}

\subsection{Per-Row Activation Counting (PRAC)}
\label{subsec:2_2_prac}

Table~\ref{tab:dram_timings} summarizes the timing parameters under both default and PRAC configurations.
PRAC is a recently introduced feature in the DDR5 JEDEC~\cite{jedec-2024-ddr5} specification that prevents read disturbance errors by providing every row with the ability to count the number of \texttt{ACT}s. 
In PRAC-enabled DRAM, each \precmd triggers an internal read–modify–write to update the activation counter for the closed row, leading to changes in several key timing parameters.
\tRAS, \tWR, and \tRTP are reduced by 16\,ns, 20\,ns, and 2.5\,ns, respectively, while \tRP and \tRC are increased by 20\,ns and 4\,ns.

A longer \tRP primarily impacts performance among the timing parameters modified by PRAC.
Compared with the default system (Figure~\ref{fig:command_sequence}(a)), the PRAC-enabled one (Figure~\ref{fig:command_sequence}(b)) incurs a longer precharge latency as the longer \tRP adds additional latency for each row-buffer miss.
Thus, frequent row-buffer misses increase the occurrence of the \tRP penalty introduced by PRAC, amplifying the performance overhead.

\section{Motivation}
\label{sec:3_Motivation}

Recent studies have investigated 
\OHprac using memory-system simulators~\cite{ramulator2, memsim, dramsim3} (Table~\ref{tab:prior_works_transposed}). 
For example, Chronus~\cite{hpca-2025-chronus} reports that even general CPU workloads experience an average slowdown of 9.7\%, with the worst-case degradation reaching up to 13.4\%.
These studies present the performance overhead as a fundamental limitation of PRAC, thereby motivating their architectural solutions. 

However, simulator-based analyses suffer from intrinsic limitations. 
Simulators may suffer from implementation bugs and insufficient fidelity, limiting their ability to accurately model the complex behavior of real-world systems.
Further, we identified a timing misconfiguration in the PRAC implementation of Ramulator 2.0~\cite{ramulator2} used by Chronus; while the parameters whose values were increased, such as $t_{\mathrm{RP}}$ and $t_{\mathrm{RC}}$, were correctly applied, the ones whose values were reduced, such as $t_{\mathrm{RAS}}$, $t_{\mathrm{RTP}}$, and $t_{\mathrm{WR}}$, were not reflected. 
After correcting this issue, the average performance overhead dropped from over 9.7\% to 5.8\%. 

Inaccurate simulator-based analyses can lead to incorrect conclusions about PRAC's practicality and implementation feasibility, which may misguide DRAM architectural decisions and future research directions.
Therefore, this work undertakes the first real-system evaluation to quantify the performance overhead induced by PRAC timing parameters and to analyze the impact of page policies on the system with PRAC.

\begin{table}[t]
  \centering
  \caption{Experimental Setup (DPC stands for DIMM per channel.)}
  \vspace{-0.05in}
  \label{tab:experimental_setup}
  \begin{tabular}{@{}c|ccc@{}}
    \toprule
    & System-A & System-B & System-C \\
    \toprule
    \makecell{Processor\\(Architecture)}
      & \makecell{i9-12900K\\(Alder Lake)}
      & \makecell{i9-14900K\\(Raptor Lake)} 
      & \makecell{Ultra 9 285K\\(Arrow Lake)} \\
    \midrule
    Main memory 
        & \multicolumn{3}{c}{\{2 Channels, 1 DPC, 16/32GB DDR5-4800, 1R/2R$\times$8\} } \\
    \midrule
    Mainboard 
      & Z690-A (ASUS) 
      & B760M (MSI) 
      & Z890-A (ASUS) \\ 
    \midrule
    BIOS ver. 
      & M.60 
      & 1505 
      & 1801 \\
    \bottomrule
  \end{tabular}
  
  \raggedright
\end{table}

\begin{figure}[!tb]
  \center
  \includegraphics[width=0.97\columnwidth]{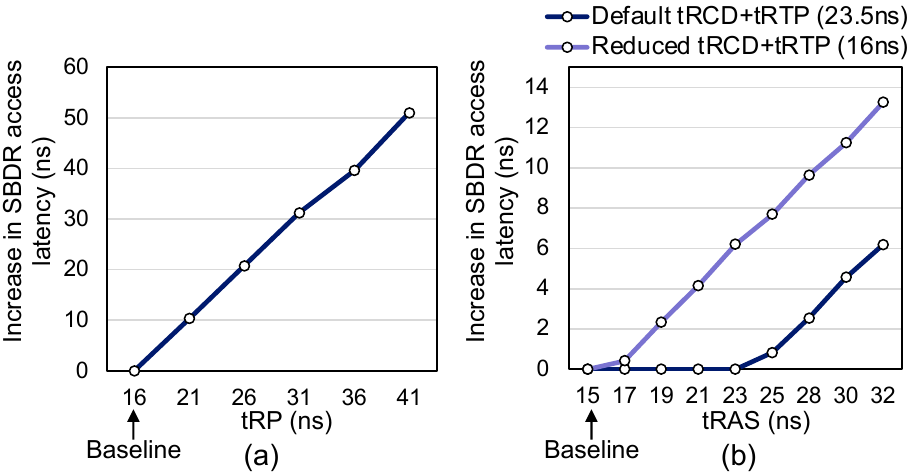}
  \vspace{-0.02in}
  \caption{A sanity check using a microbenchmark on a real-system. (a) is for tRP, and (b) is for tRAS, verifying hardware application of timing changes.}
  \label{fig:Sanity_Check}
  \vspace{-0.06in}
\end{figure}

\section{Evaluation}
\label{sec:4_Evaluation}

To accurately quantify the performance impact of the timing parameters affected by PRAC, we 1) conducted a sanity check of the modified DRAM timing parameters using a microbenchmark, 
2) measured benchmark performance under the PRAC-adjusted timing settings and identified hardware events exhibiting high correlation with \OHprac, and 3) analyzed the impact of page policy on \OHprac.
While PRAC may trigger additional refresh operations via the \texttt{ALERT\_n} signal~\cite{jedec-2024-ddr5}, when the threshold is set to 1024, the performance overhead of such refreshes is negligible under benign CPU workloads~\cite{hpca-2025-chronus,hpca-2025-autorfm}.
We tested on the latest Intel Core processors (Alder/Raptor/Arrow Lake), which support modifying DRAM timings and various features (Table~\ref{tab:experimental_setup}).\footnote{The Intel Xeon servers we could access did not permit extensive modification of DRAM timing parameters; thus, we used them only for modifying page policies in Section~\ref{subsec:4_4_RBMPKI_page_policy}.} 
All experiments were conducted on the eight performance cores (P-cores). 
We disabled Turbo Boost to ensure consistent, reliable performance measurement.
We used both single- (1R) and dual-rank (2R) $\times$8 DDR5-4800 unbuffered DIMMs configured in a 2-channel setup with one DIMM per channel (DPC).
Both \OHprac and RBMPKI were consistently lower with 2R DIMMs compared to 1R DIMMs primarily because more banks are available in the 2R case.
To ensure consistency and avoid underestimating \OHprac, we report results only for the single-rank DIMM configuration.

\begin{figure*}[tb!]
  \centering
  \includegraphics[width=0.95\textwidth]{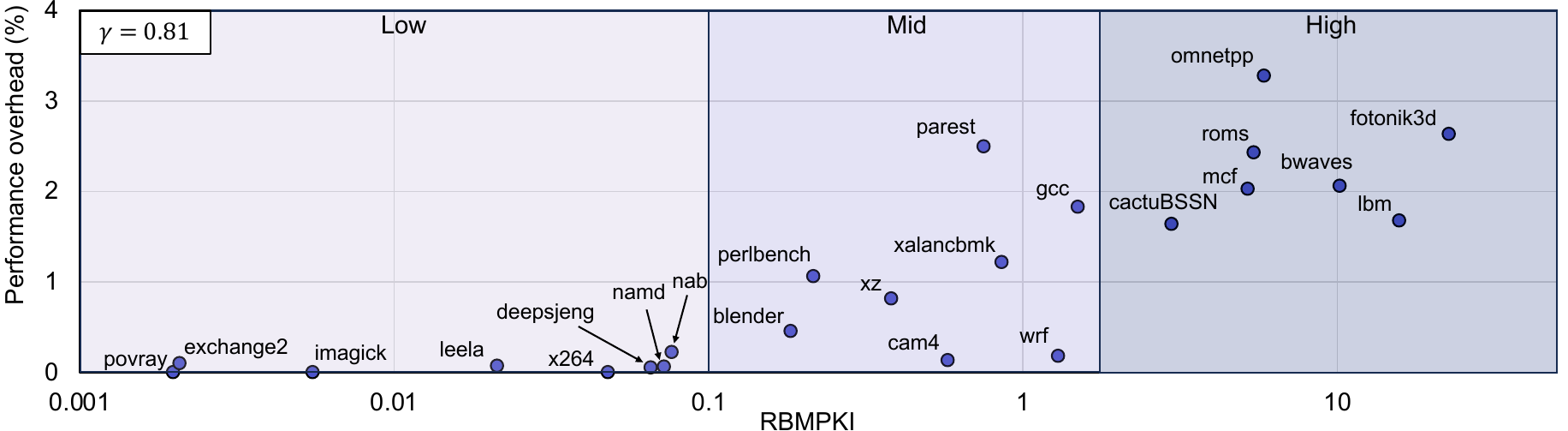}
  \vspace{-0.04in}
  \caption{\OHprac across different SPEC CPU2017 workloads~\cite{spec_cpu_2017}, categorized into \texttt{Low}, \texttt{Mid}, and \texttt{High} RBMPKI groups, with average \OHprac values of 0.06\%, 1.03\%, and 2.25\%, respectively. A strong positive correlation ($\gamma$ = 0.81) between RBMPKI and \OHprac is observed.}
  \vspace{-0.03in}
  \label{fig:RBMPKI}
\end{figure*}

\subsection{Sanity Check of PRAC-Modified DRAM Timing Parameters}
\label{subsec:4_1_sanity_check}

We modified each of the boldfaced timing parameters in Table~\ref{tab:dram_timings} via the BIOS interface, adjusting them in units of DRAM cycles.
To ensure that these changes were faithfully applied in hardware, we measured memory access latency using a microbenchmark.
It repeatedly reads two different rows in the same DRAM bank (SBDR).
%
To clearly observe the timing changes, we disabled all hardware prefetchers and flushed the target addresses from the cache using \texttt{CLFLUSH} instruction per iteration. 
We iterated sufficiently to minimize measurement noise and reported the median latency. 
Each parameter change 
was verified independently in this manner.

We conducted a sanity check using our microbenchmark to verify that the changes in timing parameter values were correctly applied to the system.
As default DRAM timing parameters are conservatively configured, they can be substantially relaxed or tightened without hurting functional correctness~\cite{hpca-2015-adaptive}.
Figure~\ref{fig:Sanity_Check} depicts the measured increase in memory access latency on \texttt{System-A} as each DRAM timing parameter is adjusted independently. 
Figure~\ref{fig:Sanity_Check}(a) illustrates the impact of changing \tRP, plotted relative to a 16\,ns baseline.  
Each pair of SBDR access in the microbenchmark involves two back-to-back row activations, incurring two \texttt{PRE}s.
Thus, an increase in \tRP leads to twice the increase in memory access latency (e.g.,  increasing \tRP from 16\,ns to 21\,ns results in a 10\,ns increase in the latency).

In the case of \tRAS (Figure~\ref{fig:Sanity_Check}(b)), reducing only \tRAS does not lower the observed latency below $t_{\mathrm{RCD}} + t_{\mathrm{RTP}}$.
Between two consecutive SBDR accesses, only the first accessed row is precharged after \tRAS, incurring a single \tRAS delay.
In the default setting, $t_{\mathrm{RCD}} + t_{\mathrm{RTP}}$ is 23.5\,ns.
Thus, even if \tRAS is set below 23.5\,ns, no change in latency is observed. 
To reduce \tRAS to the PRAC setting of 16\,ns, we first set $t_{\mathrm{RCD}} + t_{\mathrm{RTP}}$ to their minimum possible sum of 16\,ns. 
Under this configuration, access latency decreases linearly as \tRAS is reduced until 16\,ns.
Similar verification confirmed all BIOS-configured timing changes were correctly applied and maintained full functional integrity for all system workloads.

\subsection{Evaluating performance under modified DRAM timings}
\label{subsec:4_2_evaluation}

We measured performance on SPEC CPU2017~\cite{spec_cpu_2017} using the rate mode with eight P-cores and weighted speedup metric.
All 23 workloads are classified into three groups (\texttt{Low}, \texttt{Mid}, and \texttt{High}) based on their \emph{Row-Buffer Misses Per Kilo Instructions} (RBMPKI)~\cite{hpca-2025-chronus,hpca-2021-blockhammer,hpca-2024-comet}.
To compute RBMPKI, we use \texttt{perf} on Linux to measure the number of retired instructions (\texttt{instructions}) and the number of \texttt{PRE} triggered due to \emph{row-buffer miss} (\texttt{unc\_m\_pre\_count\_page\_miss}).  
The seven workloads with the highest RBMPKI are labeled as the \texttt{High} group, the next eight as \texttt{Mid}, and the remaining eight as \texttt{Low}. 
In System-A, the average RBMPKI of the \texttt{Low}, \texttt{Mid}, and \texttt{High} groups was 0.02, 0.57, and 7.82, respectively.

\OHprac on real-systems was notably lower than previously reported in simulator-based studies.
Figure~\ref{fig:RBMPKI} shows \OHprac across workloads on System-A, grouped by RBMPKI, with the hardware prefetcher enabled.
With the hardware prefetcher disabled (enabled), average overheads for \texttt{Low}, \texttt{Mid}, and \texttt{High} RBMPKI groups were 0.05\% (0.06\%), 1.34\% (1.03\%), and 2.75\% (2.25\%), respectively.  
The overall average \OHprac was 1.33\% (1.06\%), with a peak overhead of 3.77\% (3.28\%). 
Compared to the simulator-based results~\cite{hpca-2025-chronus}, this corresponds to a reduction of up to 7.29$\times$ (9.15$\times$) in average overhead and 3.55$\times$ (4.09$\times$) in the worst case.

We observed that \OHprac increases with higher RBMPKI.
Statistical analysis yields a Pearson correlation coefficient of 0.81 between \OHprac and the logarithm of RBMPKI.
With a $p$-value ($<10^{-5}$), this indicates a significant and strong positive correlation between RBMPKI and \OHprac.

\subsection{RBMPKI with different page policy}
\label{subsec:4_4_RBMPKI_page_policy}

We observed an evident correlation between \OHprac and workload's RBMPKI.
As the memory controller’s page policy plays a significant role in determining RBMPKI by controlling the row activation and precharge behavior,
we analyzed how variations in page policy affect RBMPKI.
Furthermore, we investigate which page policy can most effectively minimize \OHprac by reducing RBMPKI.

As the CPUs listed in Table~\ref{tab:experimental_setup} do not expose interfaces for inspecting or modifying the page policy, we conducted our experiments on an Intel Xeon Platinum 8260 (Cascade Lake) processor.  
By reading the \texttt{idletime} register~\cite{2014-intel-idletime}, we confirmed that its default page policy is \emph{adaptive}.
We selected four workloads from each previously defined RBMPKI group (\texttt{Low}, \texttt{Mid}, and \texttt{High}).
As different page policies may use different address mappings~\cite{intel-2024-pagepolicy}, complicating direct comparisons, we emulated \emph{open} and \emph{close} behavior by tuning the parameters of the \texttt{idletime} register.

\begin{table}[tb!]
  \caption{Parameter settings for page policies. (\textsuperscript{\dag}, \textsuperscript{*}) stand for setting the parameters to the (maximum, minimum) values.}
  \vspace{-0.05in}
  \label{tab:page_policy_params}
  \centering
  \scriptsize
  \renewcommand{\arraystretch}{1.1}
  \setlength{\tabcolsep}{1pt}
  \begin{tabularx}{\columnwidth}{@{}l
    >{\centering\arraybackslash}X
    >{\centering\arraybackslash}X
    >{\centering\arraybackslash}X@{}}
    \toprule
    Parameter & \makecell{\emph{open}} & \makecell{\emph{adaptive}} & \makecell{\emph{close}} \\
    \midrule
    \texttt{Win\_size}            & 255\textsuperscript{\dag}  & 64   & 1   \\
    \texttt{IDLE\_PAGE\_RST\_VAL} (cycle) & 63\textsuperscript{\dag}   & 8    & 0\textsuperscript{*}   \\
    \texttt{OPC\_TH}              & 127\textsuperscript{\dag}  & 6    & 0\textsuperscript{*}   \\
    \texttt{PPC\_TH}              & 0\textsuperscript{*}       & 6    & 127\textsuperscript{\dag} \\
    \bottomrule
  \end{tabularx}
  \vspace{-0.1in}
\end{table}

The following four parameters govern our emulation:

\begin{itemize}
  \item \texttt{Win\_size}: The number of memory requests to track in the recent access window.
  \item \texttt{IDLE\_PAGE\_RST\_VAL}: The delay before issuing a \texttt{PRE} after an \texttt{ACT} when no further access occurs.
  \item \texttt{OPC\_TH} (Overdue Page Close Threshold): The threshold for row-buffer misses; if exceeded, the controller shortens \texttt{IDLE\_PAGE\_RST\_VAL} to close pages more aggressively.
  \item \texttt{PPC\_TH} (Premature Page Close Threshold): The threshold for premature closes that disrupt row-buffer hits; if exceeded, the controller increases \texttt{IDLE\_PAGE\_RST\_VAL} to keep pages open longer.
\end{itemize}
\vspace{-0.05in}

\medskip

We compared three-page policies, \emph{open}, \emph{adaptive open} (default), and \emph{close}, using the parameter settings detailed in Table~\ref{tab:page_policy_params}.
\emph{Close} achieved the lowest row-buffer miss ratio on average among the three policies (see Figure~\ref{fig:policy_RBMPKI_IPC})
due to its proactive precharge mechanism, which closes the activated row immediately after each access, thereby returning the bank to the row-buffer empty state more frequently and reducing the likelihood of row-buffer misses.
In contrast, \emph{open} exhibits the highest row-buffer hit and miss rates, as it keeps rows open to maximize the row-buffer hit ratio.
\emph{Adaptive} operates in between: it behaves similarly to \emph{open} for workloads with low RBMPKI and shifts toward  \emph{close} as RBMPKI increases.

We further evaluated the IPC of workloads under different page policies.
Compared to \emph{open}, \emph{adaptive}, and \emph{close} achieve 1.84\% and 2.13\% higher IPC on average for the \texttt{High} group, respectively.
Notably, in the \texttt{High} group, both \emph{close} and \emph{adaptive} significantly outperform \emph{open}
due to the higher latency associated with \emph{open} on a row-buffer miss~\cite{isca-2014-decoupling}.
Unlike \emph{close}, which increases the occurrence of the row-buffer empty state to reduce misses, \emph{open} causes more row-buffer misses, leading to an increased \tRP penalty.
This additional delay increases access latency, reducing IPC. 
By avoiding it, \emph{close} not only minimizes RBMPKI but also achieves the highest IPC.  
As \emph{close} yields the highest performance gain and lowest RBMPKI, it is the most effective strategy for mitigating the performance overhead imposed by PRAC timing parameters.

\section{Conclusion}
\label{sec:5_conclusion}
\rev{
\begin{figure}[!tb]
  \centering
  \includegraphics[width=0.99\columnwidth]{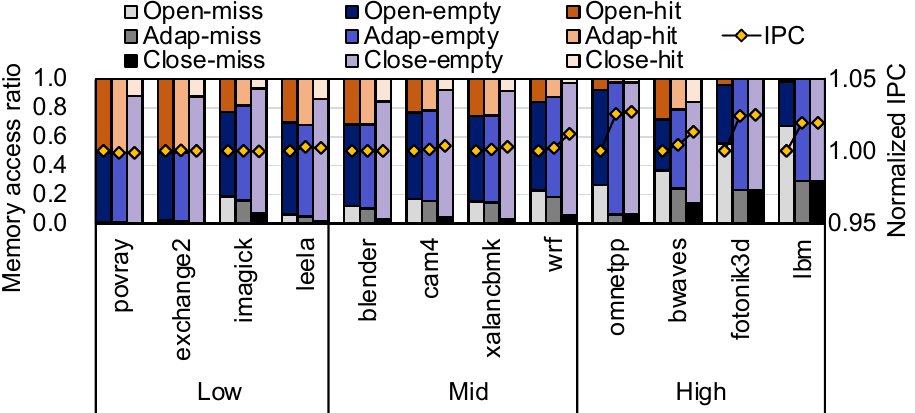}
  \caption{
  Breakdown of different memory access ratio (row-buffer hit, empty, miss) and IPC for selected workloads under \emph{open}, \emph{adaptive}, and \emph{close}. IPC is normalized to \emph{open}.}
  \vspace{-0.07in}
  \label{fig:policy_RBMPKI_IPC}
\end{figure}}

To address DRAM read disturbance errors, Per-Row Activation Counting (PRAC) has been incorporated into the latest DRAM standards.
We presented the first real-system evaluation of PRAC timing overheads.
Our key findings are: 1) PRAC's average performance overhead is only 1.06\% (maximum 3.28\%) on modern CPUs with SPEC CPU2017, up to 9.15$\times$ lower than prior simulator-based reports.
We partly attributed this discrepancy to simulator inaccuracies, including timing misconfigurations we identified.
2) The overhead correlates strongly with Row-Buffer Misses Per Kilo Instructions (RBMPKI).
3) Using a close page policy or its variants in memory controllers effectively minimizes RBMPKI and PRAC overhead, also improving performance.
These results suggest that PRAC is more practical than previously thought, especially with appropriate page policy management.

\balance
\bibliographystyle{IEEEtran}
\bibliography{ref}

\begin{thebibliography}{10}
\providecommand{\url}[1]{#1}
\csname url@samestyle\endcsname
\providecommand{\newblock}{\relax}
\providecommand{\bibinfo}[2]{#2}
\providecommand{\BIBentrySTDinterwordspacing}{\spaceskip=0pt\relax}
\providecommand{\BIBentryALTinterwordstretchfactor}{4}
\providecommand{\BIBentryALTinterwordspacing}{\spaceskip=\fontdimen2\font plus
\BIBentryALTinterwordstretchfactor\fontdimen3\font minus \fontdimen4\font\relax}
\providecommand{\BIBforeignlanguage}[2]{{%
\expandafter\ifx\csname l@#1\endcsname\relax
\typeout{** WARNING: IEEEtran.bst: No hyphenation pattern has been}%
\typeout{** loaded for the language `#1'. Using the pattern for}%
\typeout{** the default language instead.}%
\else
\language=\csname l@#1\endcsname
\fi
#2}}
\providecommand{\BIBdecl}{\relax}
\BIBdecl

\bibitem{hpca-2025-chronus}
O.~Canpolat, A.~G. Ya{\u{g}}l{\i}k{\c{c}}{\i}, G.~F. Oliveira, A.~Olgun, N.~Bostanc{\i}, I.~E. Yuksel, H.~Luo, O.~Ergin, and O.~Mutlu, ``{Chronus: Understanding and Securing the Cutting-Edge Industry Solutions to DRAM Read Disturbance},'' in \emph{HPCA}, 2025.

\bibitem{hpca-2025-autorfm}
M.~Qureshi, ``{AutoRFM: Scaling Low-Cost in-DRAM Trackers to Ultra-Low Rowhammer Thresholds},'' in \emph{HPCA}, 2025.

\bibitem{isca-2025-mopac}
S.~Vittal, S.~Qazi, P.~Das, and M.~Qureshi, ``{MoPAC: Efficiently Mitigating Rowhammer with Probabilistic Activation Counting},'' in \emph{ISCA}, 2025.

\bibitem{hpca-2025-qprac}
J.~Woo, S.~C. Lin, P.~J. Nair, A.~Jaleel, and G.~Saileshwar, ``{QPRAC: Towards Secure and Practical PRAC-based Rowhammer Mitigation using Priority Queues},'' in \emph{HPCA}, 2025.

\bibitem{chronus-arxiv-2025}
\BIBentryALTinterwordspacing
O.~Canpolat, A.~G. Yağlıkçı, G.~F. Oliveira, A.~Olgun, N.~Bostancı, İsmail Emir~Yüksel, H.~Luo, O.~Ergin, and O.~Mutlu, ``{Chronus: Understanding and Securing the Cutting-Edge Industry Solutions to DRAM Read Disturbance},'' 2025. [Online]. Available: \url{https://arxiv.org/abs/2502.12650}
\BIBentrySTDinterwordspacing

\bibitem{qprac-ae}
J.~Woo, S.~C. Lin, P.~J. Nair, A.~Jaleel, and G.~Saileshwar, ``{QPRAC (HPCA 2025)},'' \url{https://github.com/sith-lab/qprac/blob/main/perf_analysis/src/dram/impl/DDR5-PRAC.cpp}, 2025 (accessed Oct 23rd, 2025).

\bibitem{micro-2024-mess}
P.~Esmaili-Dokht, F.~Sgherzi, V.~S. Girelli, I.~Boixaderas, M.~Carmin, A.~Monemi, A.~Armejach, E.~Mercadal, G.~Llort, P.~Radojkovi{\'c} \emph{et~al.}, ``{A Mess of Memory System Benchmarking, Simulation and Application Profiling},'' in \emph{MICRO}, 2024.

\bibitem{luo2025cleaningmess}
\BIBentryALTinterwordspacing
H.~Luo, A.~Olgun, M.~Makeenkova, F.~N. Bostanci, G.~F. Oliveira, A.~G. Yaglikci, and O.~Mutlu, ``{Cleaning up the Mess},'' 2025. [Online]. Available: \url{https://arxiv.org/abs/2510.15744}
\BIBentrySTDinterwordspacing

\bibitem{taco-prefetch}
J.~Lee, H.~Kim, and R.~Vuduc, ``{When Prefetching Works, When It Doesn't, and Why},'' \emph{ACM Trans. Archit. Code Optim.}, vol.~9, no.~1, Mar. 2012.

\bibitem{spec_cpu_2017}
S.~P.~E. Corp., ``{SPEC CPU2017},'' \url{https://www.spec.org/cpu2017}.

\bibitem{intel-2024-pagepolicy}
Intel, ``{Performance Differences for Open-Page / Close-Page Policy},'' \url{https://www.intel.com/content/www/us/en/content-details/826015/performance-differences-for-open-page-close-page-policy.html}, 2024.

\bibitem{ramulator2}
H.~Luo, Y.~C. Tu{\u{g}}rul, F.~N. Bostanc{\i}, A.~Olgun, A.~G. Ya{\u{g}}l{\i}k{\c{c}}{\i}, and O.~Mutlu, ``{Ramulator 2.0: A Modern, Modular, and Extensible DRAM Simulator},'' in \emph{IEEE Computer Architecture Letters}, 2023.

\bibitem{memsim}
``{Memsim: Memory System Simulator},'' \url{https://github.com/mqureshi4/memsim}, 2024.

\bibitem{dramsim3}
S.~Li, Z.~Yang, D.~Reddy, A.~Srivastava, and B.~Jacob, ``{DRAMsim3: A Cycle-Accurate, Thermal-Capable DRAM Simulator},'' in \emph{IEEE Computer Architecture Letters}, 2020.

\bibitem{jedec-2024-ddr5}
JEDEC, ``{DDR5 SDRAM Standard},'' 2024.

\bibitem{hpca-2015-adaptive}
D.~Lee, Y.~Kim, G.~Pekhimenko, S.~Khan, V.~Seshadri, K.~Chang, and O.~Mutlu, ``{Adaptive-latency DRAM: Optimizing DRAM timing for the common-case},'' in \emph{HPCA}, 2015.

\bibitem{hpca-2021-blockhammer}
A.~G. Yağlikçi, M.~Patel, J.~S. Kim, R.~Azizi, A.~Olgun, L.~Orosa, H.~Hassan, J.~Park, K.~Kanellopoulos, T.~Shahroodi, S.~Ghose, and O.~Mutlu, ``{BlockHammer: Preventing RowHammer at Low Cost by Blacklisting Rapidly-Accessed DRAM Rows},'' in \emph{HPCA}, 2021.

\bibitem{hpca-2024-comet}
F.~N. Bostanci, I.~E. Y{\"u}ksel, A.~Olgun, K.~Kanellopoulos, Y.~C. Tu{\u{g}}rul, A.~G. Ya{\u{g}}li{\c{c}}i, M.~Sadrosadati, and O.~Mutlu, ``{CoMeT: Count-Min-Sketch-based Row Tracking to Mitigate RowHammer at Low Cost},'' in \emph{HPCA}, 2024.

\bibitem{2014-intel-idletime}
I.~Corporation, ``{Intel® Xeon Processor E5 v2 Product Family Datasheet, Volume 2},'' 2014.

\bibitem{isca-2014-decoupling}
S.~O, Y.~H. Son, N.~S. Kim, and J.~Ahn, ``{Row-buffer Decoupling: A Case for Low-latency DRAM Microarchitecture},'' in \emph{ISCA}, 2014.

\end{thebibliography}

\end{document}